\documentclass[twocolumn,floatfix,superscriptaddress,a4paper,showkeys,nofootinbib,reprint]{revtex4-1}

\usepackage[colorlinks=true,linktocpage=true,linkcolor=blue,citecolor=blue,allcolors=blue]{hyperref}
\usepackage{epsfig}
\usepackage{latexsym}
\usepackage[utf8]{inputenc}
\usepackage{xspace}
\usepackage{indentfirst}
\usepackage{enumitem}
\usepackage{color}

\usepackage{setspace}% http://ctan.org/pkg/setspace
\usepackage{lipsum}% http://ctan.org/pkg/lipsum

\usepackage{todonotes}

\usepackage{amsmath}
\usepackage{amssymb}
\usepackage[english]{babel}
\usepackage{url}

\newcommand{\nBB}{n_{B(\bar{B})}}

\newcommand{\eq}[1]{\begin{align} #1 \end{align}}

\begin{document}

\title{
Traces of the nuclear liquid-gas phase transition in the analytic properties of hot QCD
}

\author{Oleh Savchuk}
\affiliation{Physics Department, Taras Shevchenko National University of Kyiv, 03022 Kyiv, Ukraine}
\author{Volodymyr Vovchenko}
\affiliation{Institut f\"ur Theoretische Physik,
Goethe Universit\"at Frankfurt, D-60438 Frankfurt am Main, Germany}
\affiliation{Frankfurt Institute for Advanced Studies, Giersch Science Center,
D-60438 Frankfurt am Main, Germany}
\author{Roman V. Poberezhnyuk}
\affiliation{Bogolyubov Institute for Theoretical Physics, 03680 Kyiv, Ukraine}
\affiliation{Frankfurt Institute for Advanced Studies, Giersch Science Center, D-60438 Frankfurt am Main, Germany}
\author{Mark I. Gorenstein}
\affiliation{Bogolyubov Institute for Theoretical Physics, 03680 Kyiv, Ukraine}
\affiliation{Frankfurt Institute for Advanced Studies, Giersch Science Center, D-60438 Frankfurt am Main, Germany}
\author{Horst Stoecker}
\affiliation{Institut f\"ur Theoretische Physik,
Goethe Universit\"at Frankfurt, D-60438 Frankfurt am Main, Germany}
\affiliation{Frankfurt Institute for Advanced Studies, Giersch Science Center,
D-60438 Frankfurt am Main, Germany}
\affiliation{GSI Helmholtzzentrum f\"ur Schwerionenforschung GmbH, D-64291 Darmstadt, Germany}

\date{\today}

\begin{abstract}
The nuclear liquid-gas transition at normal nuclear densities, $n \sim n_0 = 0.16$~fm$^{-3}$, and small temperatures, $T \sim 20$~MeV, has a large influence on analytic properties of the QCD grand-canonical thermodynamic potential.
A classical van der Waals equation is used to determine these unexpected features due to dense cold matter qualitatively.
The existence of the nuclear matter critical point results in thermodynamic branch points, which are located at complex chemical potential values, for $T > T_c \simeq 20$~MeV, and exhibit a moderate model dependence up to rather large temperatures $T \lesssim 100$~MeV.
The behavior at higher temperatures is studied using the van der Waals hadron resonance gas~(vdW-HRG) model. 
The baryon-baryon interactions have a decisive influence on 
the QCD thermodynamics close to $\mu_B = 0$.
In particular, nuclear matter singularities limit the radius of convergence $r_{\mu_B/T}$ of the Taylor expansion in $\mu_B/T$, with $r_{\mu_B/T} \sim 2-3$ values at $T \sim 140-170$~MeV obtained in the vdW-HRG model. 
\end{abstract}
% \pacs{15.75.Ag, 24.10.Pq}


\maketitle

%\begin{spacing}{1.9}

\section{Introduction}

The thermodynamic properties of QCD  at finite temperatures and densities are important issues of modern high-energy nuclear physics.
Of particular interest are the phase structure of QCD matter and the nature of the hadron-parton transition.
At zero baryon density, i.e. at $\mu_B = 0$, this transition is a crossover, according to lattice QCD simulations~\cite{Aoki:2006we}.
The nature of this transition at finite densities is not established yet.
The experimental search for the hypothetical QCD chiral critical point~(CP)~\cite{Stephanov:1998dy} is performed at non-zero intermediate baryon densities using measurements of fluctuations in heavy-ion collisions~\cite{Stephanov:2008qz,Koch:2008ia,Gazdzicki:2015ska,Luo:2017faz} as well as indirect lattice gauge theory methods, such as a Taylor expansion around $\mu_B = 0$~\cite{Allton:2002zi,Gavai:2008zr} or analytic continuation from imaginary $\mu_B$~\cite{deForcrand:2002hgr,DElia:2002tig}.
Current high quality lattice QCD data at physical quark masses show no evidence or signatures of a chiral CP and disfavor the existence of a phase transition of first or second order at moderate baryon densities $\mu_B/T \lesssim \pi$~\cite{deForcrand:2006pv,Bazavov:2017dus,Vovchenko:2017gkg,Fodor:2018wul}.
The location or even the existence of that CP is not settled to date.
Possibilities for a phase transition at large baryon densities can be explored in heavy-ion collisions at moderate collision energies, such as the CBM experiment at FAIR~\cite{Ablyazimov:2017guv}, or through precision neutron star merger observations and gravitational wave astronomy~\cite{Most:2018eaw,Bauswein:2018bma}.

In contrast to the chiral QCD CP, an existence of the nuclear liquid-gas phase transition with an associated CP at $T_c \simeq 20$~MeV and $\mu_B^c \simeq 900$~MeV 
is better established both
theoretically~\cite{Sauer:1976zzf,Csernai:1986qf,Serot:1984ey,Zimanyi:1990np,Brockmann:1990cn} and experimentally~\cite{Pochodzalla:1995xy,Natowitz:2002nw,Karnaukhov:2003vp}~(see Ref.~\cite{Elliott:2013pna} for the current empirical estimates of the CP location).
This transition is also accessible on the lattice through an effective theory~\cite{Fromm:2012eb}.
Recently it has been pointed out that nuclear matter criticality has a sizeable influence on conserved charges susceptibilities in hot QCD matter, both in the vicinity of the crossover temperature region at $\mu_B = 0$~\cite{Vovchenko:2016rkn}, and along the phenomenological freeze-out curve in heavy-ion collisions~\cite{Fukushima:2014lfa,Vovchenko:2017ayq,Poberezhnyuk:2019pxs}.
In spite of the fact that the nuclear phase transition ends at $T_c \sim 20$~MeV, its remnants appear to survive in certain observables to much higher temperatures.

A presence of a phase transition and a CP is imprinted in analytic properties of a thermodynamic potential.
The pressure function, in particular, becomes a multi-valued function of the chemical potential, and exhibits branch cut singularities~\cite{Stephanov:2006dn}.
At subcritical temperatures these singularities correspond to spinodal instabilities, at $T = T_c$ the singularities merge at the CP, and at $T > T_c$ the singularities lie at complex values of the chemical potential~\cite{Vovchenko:2019hbc}.
Phase transitions are smoothed out in a finite volume, their remnants are characterized there by the Lee-Yang zeroes of the grand partition function~\cite{Yang:1952be,Lee:1952ig}.

The thermodynamic branch points associated with the nuclear liquid-gas transition are studied in detail in the present work.
First, analytic results on the basis of the classical van der Waals~(vdW) equation are presented in Sec.~\ref{sec:vdwcl}.
These are compared at intermediate temperatures~($T \lesssim 100$~MeV) with numerical results obtained using quantum vdW, Walecka, and Skyrme models of nuclear matter~(Sec.~\ref{sec:num}).
An extrapolation to higher temperatures is achieved in the framework of the vdW-HRG model, with a focus on the influence of the nuclear matter LGPT singularities on convergence properties of the Taylor expansion in $\mu_B/T$ around $\mu_B=0$~(Sec.~\ref{sec:vdWHRG}).
Summary in Sec.~\ref{sec:summary} closes the article.

\section{Thermodynamic branch points of a liquid-gas phase transition}
\label{sec:vdwcl}

Let us first consider the system of interacting nucleons as a classical real gas described by the vdW equation. The pressure reads~\cite{GNS}
\eq{\label{eq:pvdwcl}
p(T,n) = \frac{T \, n}{1 - b \, n} - a \, n^2,
}
where $a > 0$ and $b > 0$ correspond, respectively, to attractive and repulsive interactions.
In the grand-canonical ensemble~(GCE) the particle number density~$n(T,\mu)$ is defined by a transcendental equation~\cite{Vovchenko:2015xja}:
\eq{\label{eq:eqgce}
e^{\mu/T} = \frac{n}{\phi(T) \, (1-bn)} \, \exp\left[ \frac{bn}{1-bn} - \frac{2an}{T} \right]~.
}
Here
\eq{
\phi(T) & = \frac{d \, m^2 \, T}{2 \pi^2} \, K_2(m/T),
}
where $d$ is the degeneracy factor and $m$ is particle's mass\footnote{In our consideration $d=4$, $m=938$~MeV for nucleons. We neglect the small difference between proton and neutron masses.}.
Substituting $n(T,\mu)$ into Eq.~\eqref{eq:pvdwcl} then allows one to reconstruct the GCE pressure function, i.e. a full thermodynamic potential in the GCE.

At given values of $T$ and (complex)~$\mu$, Eq.~\eqref{eq:eqgce} may have more than a single solution, meaning that $n(T,\mu)$ is a multi-valued function.
This multivalueness entails an existence of branch points. Early studies of the branch points for the classical vdW equation can be found in Ref.~\cite{hemmer1964yang}.
Here we present a systematic analysis of the behavior of branch points related to the nuclear liquid-gas transition and their relevance for the QCD phase diagram.

The branch points of $n(T, \mu)$ are defined through the equation \cite{Stephanov:2006dn}
\eq{\label{eq:br}
(\partial \mu / \partial n)_T = 0~.
}
Applied to Eq.~\eqref{eq:eqgce} this yields
\eq{\label{eq:nbr}
\frac{2 a n_{\rm br}}{T} \, (1 - b n_{\rm br})^2 = 1.
}
Equation~\eqref{eq:nbr} is a cubic equation for $n_{\rm br}$ defining the branch points. $\mu_{\rm br}$ is recovered by substituting $n_{\rm br}$ into Eq.~\eqref{eq:eqgce}.

The cubic equation~\eqref{eq:nbr} has three roots which are explicitly obtained using Cardano's formulas:
\eq{\label{eq:firstroot}
&n_{\rm br1,2}=\frac{1}{b}\left(-\frac{q_1+q_2}{2}\pm i \sqrt{3}\frac{q_1-q_2}{2}+\frac{2}{3}\right),\\
 &n_{\rm br3}=\frac{1}{b}\left(q_1+q_2+\frac{2}{3}\right),\label{eq:nbr3}
}
where $q_{1,2}=\sqrt[3]{A\pm\sqrt{\Delta}}$ with 
\eq{
A=\frac{1}{108 }\left(\frac{27 b}{a}T-4\right),~~~~~\Delta=A^2-\frac{1}{9^3}.
}

The third root, $n_{\rm br3}$, given by Eq.~(\ref{eq:nbr3}), is real at all values of $T$ and is larger than the %dense packing limit
limiting density
of vdW excluded volume: $n_{\rm br3} > 1/b$.
Therefore, $n_{\rm br3}$ is not accessible in the region of physical solutions at any temperature and does not appear to be connected to the existence of the first-order phase transition in the vdW equation. $n_{\rm br3}$ will thus be omitted from consideration in the following.

The behavior of the two relevant roots (\ref{eq:firstroot}) depends qualitatively on the value of the temperature.
The two roots are real at subcritical temperatures, 
$\Delta<0\Leftrightarrow T <~T_c=8a/(27b)$.
They correspond to the spinodal points of the subcritical isotherms, i.e. $(\partial p / \partial n)_T = 0$.

At the critical temperature, $\Delta=0$, $T = T_c$, the two roots become degenerate.
They coincide with the CP location, $n_{\rm br1} = n_{\rm br2} = n_c = 1/(3b)$.

At supercritical temperatures, $\Delta>0\Leftrightarrow T > T_c$, the two roots correspond to a pair of complex conjugate numbers, i.e. the singularities lie in the complex plane.
This is a manifestation of the so-called crossover transition~\cite{Stephanov:2006dn}.

In the present work we do only consider the analytic properties of pure phases and will not consider Maxwell's mixed phase construction.

\section{Branch points of the nuclear liquid-gas transition}
\label{sec:num}

In this section temperature dependence of the location of branch points associated with the nuclear liquid-gas transition is evaluated using different models of nuclear matter.
The list of models considered is given below.

%\subsection{Classical van der Waals}
\subsection{van der Waals}

As the simplest model for the nuclear liquid-gas transition we take the classical van der Waals equation~\eqref{eq:pvdwcl} for nucleons. 
We take the vdW parameter values $a = 329$~MeV~fm$^3$ and $b = 3.42$~fm$^3$ from Ref.~\cite{Vovchenko:2015vxa}.
These parameter values yield the binding energy of 16~MeV in the nuclear ground state at $n = n_0 = 0.16$~fm$^{-3}$ in the vdW model extended to include the Fermi statistics (see below). 
The locations of branch points are evaluated using Eqs.~\eqref{eq:eqgce} and \eqref{eq:nbr}.
The classical vdW equation predicts a nuclear liquid-gas transition with a CP at the following location:
\eq{
T_c = \frac{8a}{27b} \simeq 28.5~\text{MeV}, \quad n_c = \frac{1}{3b} \simeq 0.10~\text{fm}^{-3}.
}
The model captures the qualitative features associated with a first-order phase transition but it is not accurate at small temperatures, where Fermi-Dirac statistics cannot be neglected.
For the same reasons, the classical vdW equation overestimates the value of the critical temperature by about 10~MeV~\cite{Vovchenko:2015vxa,Fedotkin:2019bhq}.
% The temperature dependencies of real and imaginary parts of $\mu_{\rm br}$ coordinates are depicted in Fig.~\ref{fig:NMcompare}.

%\subsection{Quantum van der Waals}
%\label{subsec:qstat}

The quantum statistical effects are taken into account in the \emph{quantum van der Waals} model~(QvdW)~\cite{Vovchenko:2015vxa}.
The QvdW model is defined by the following equations:
\eq{
\label{eq:pqvdw}
p(T,\mu) & = p_{\rm id}(T, \mu^*) - a n^2, \\
\label{eq:nqvdw}
n(T,\mu) & = (1 - b \, n ) \, n_{\rm id}(T, \mu^*), \\
\label{eq:mustqvdw}
\mu^* & = \mu - b \, p_{\rm id}(T, \mu^*) + 2 \, a \, n.
}
Here $n_{\rm id},~p_{\rm id}$ are, respectively, the density and the pressure of the ideal Fermi gas.
In the Boltzmann approximation Eqs.~(\ref{eq:pqvdw})-(\ref{eq:mustqvdw}) reduce to the classical vdW equations~(\ref{eq:pvdwcl}) and~(\ref{eq:eqgce}).

Thermodynamic functions at fixed $T$ and $\mu$ are usually determined by solving Eq.~\eqref{eq:mustqvdw} numerically with respect to (w.r.t.) $\mu^*$, which then allows calculating all other quantities.
The vdW parameter values are the same as for the classical vdW model above.
The QvdW model predicts a CP at $T_c \simeq 19.7$~MeV and $n_c \simeq 0.07$~fm$^{-3}$~($\mu_c \simeq 908$~MeV).

One needs to evaluate the derivative $(\partial \mu / \partial n)_T$ in order to determine the thermodynamic branch points. To do that we apply the derivative w.r.t. $n$ at fixed $T$ to Eqs.~\eqref{eq:nqvdw} and~\eqref{eq:mustqvdw}, which allows to determine $(\partial \mu / \partial n)_T$ explicitly. The resulting equation $(\partial \mu / \partial n)_T = 0$ for the branch points reads
\eq{\label{eq:qvdwbr}
\frac{2 \, a \, n_{\rm br}}{T} \, (1 - b n_{\rm br})^2 \, \omega_{\rm id}(T, \mu_{\rm br}^*) = 1.
}
Here $\omega_{\rm id}(T, \mu^*)$ is the scaled variance of particle number fluctuations of an ideal quantum gas in the GCE:
\eq{\label{eq:wid}
\omega_{\rm id} (T, \mu^*) & = 1 -
\frac{d\,\eta}{2\,\pi^2\,n_{\rm id}} \int_0^{\infty} dk k^2
\nonumber
\\
& \quad \times\left[ \exp\left(\frac{\sqrt{m^2+k^2}-\mu^*}{T}\right) + \eta\right]^{-2},
}
with $\eta = +1$ for fermions.
In the Maxwell-Boltzmann approximation~($\eta = 0$) one has $\omega_{\rm id} = 1$ and Eq.~\eqref{eq:qvdwbr} reduces to Eq.~\eqref{eq:nbr} of the classical vdW model.

\begin{figure}[t]
  \centering
  \includegraphics[width=.48\textwidth]{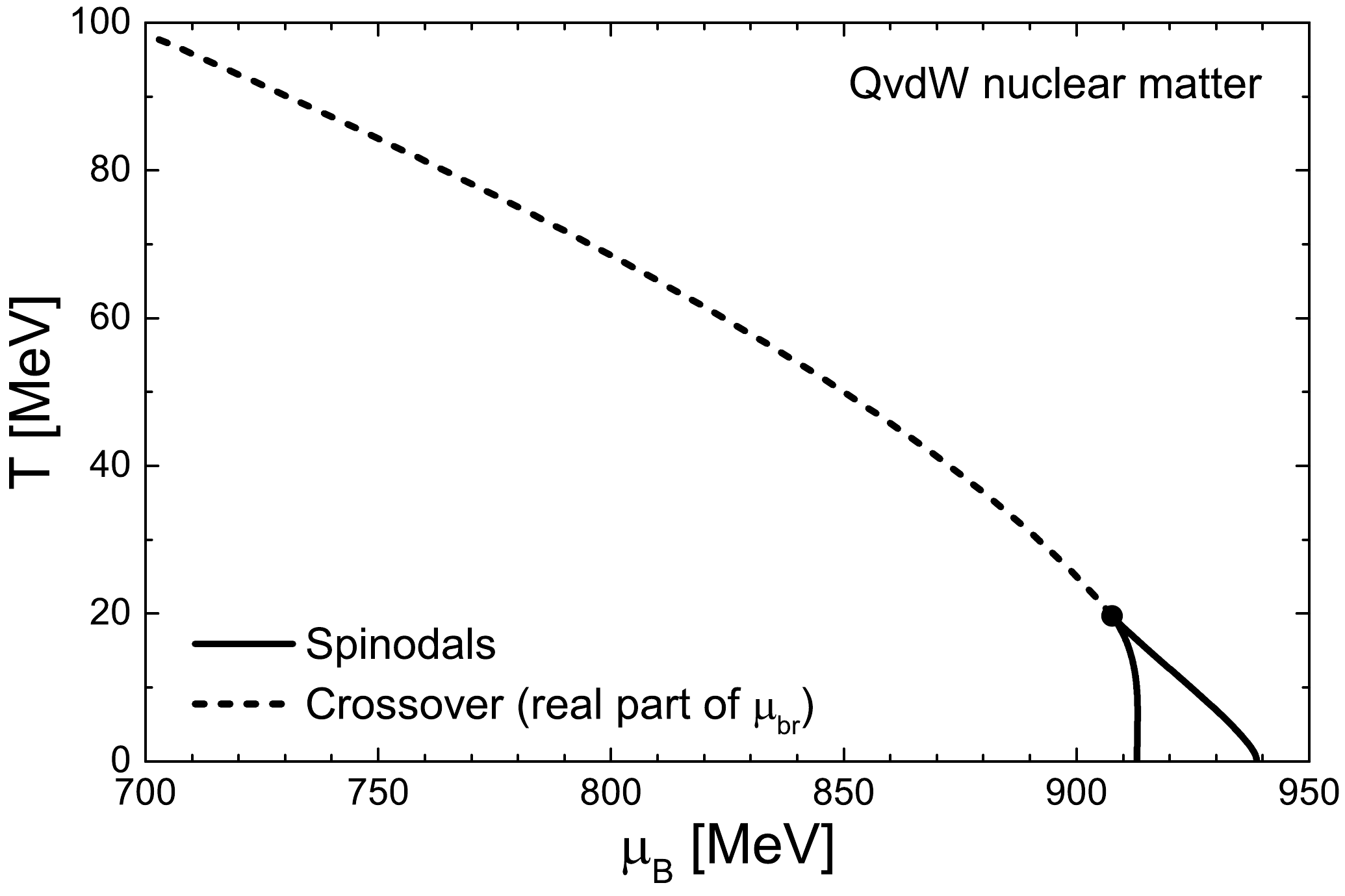}
  \caption{
   Locations of thermodynamic branch points associated with the nuclear liquid-gas transition in the $\mu_B$-$T$ plane evaluated within the quantum van der Waals model of nuclear matter.
   The two solid lines depict the two spinodals of the nuclear liquid-gas transition at $T < T_c$.
   The dashed line corresponds to the real part of the crossover branch point $\mu_{\rm br}$ at $T > T_c$.
   The circle 
   %corresponds to 
   represents
   the CP.
  }
  \label{fig:NMQvdW}
\end{figure}

Here we solve Eq.~\eqref{eq:qvdwbr} numerically to determine $\mu^*_{\rm br}$\footnote{$n_{\rm br}$ is calculated at a given $\mu^*_{\rm br}$ from Eq.~\eqref{eq:nqvdw}}.
At $T = T_c$ the solution of Eq.~\eqref{eq:qvdwbr} corresponds to the CP.
We use the CP as a starting point of the numerical procedure and move in small steps in temperature independently for $T > T_c$~(crossover) and $T < T_c$~(first-order phase transition), using the solution at the previous step as an initial guess for the next one.

Figure~\ref{fig:NMQvdW} depicts the resulting chemical potential values corresponding to the branch points. At $T<T_c$ there are two real solutions which correspond to the spinodals of the first-order phase transition, as discussed in Sec.~\ref{sec:vdwcl} for the classical vdW equation. These are depicted in Fig.~\ref{fig:NMQvdW} by two solid lines.
At $T = T_c$ the two roots become degenerate at the CP.
At $T > T_c$, $\mu_{\rm br}$ have non-zero imaginary part, the branch points correspond to two complex conjugate roots.
The behavior of the real part $\mu_{\rm br}^R \equiv \text{Re} \, [\mu_{\rm br}]$ at $T > T_c$ is shown in Fig.~\ref{fig:NMQvdW} by the dashed line.

\begin{figure*}[t]
  \centering
  \includegraphics[width=.49\textwidth]{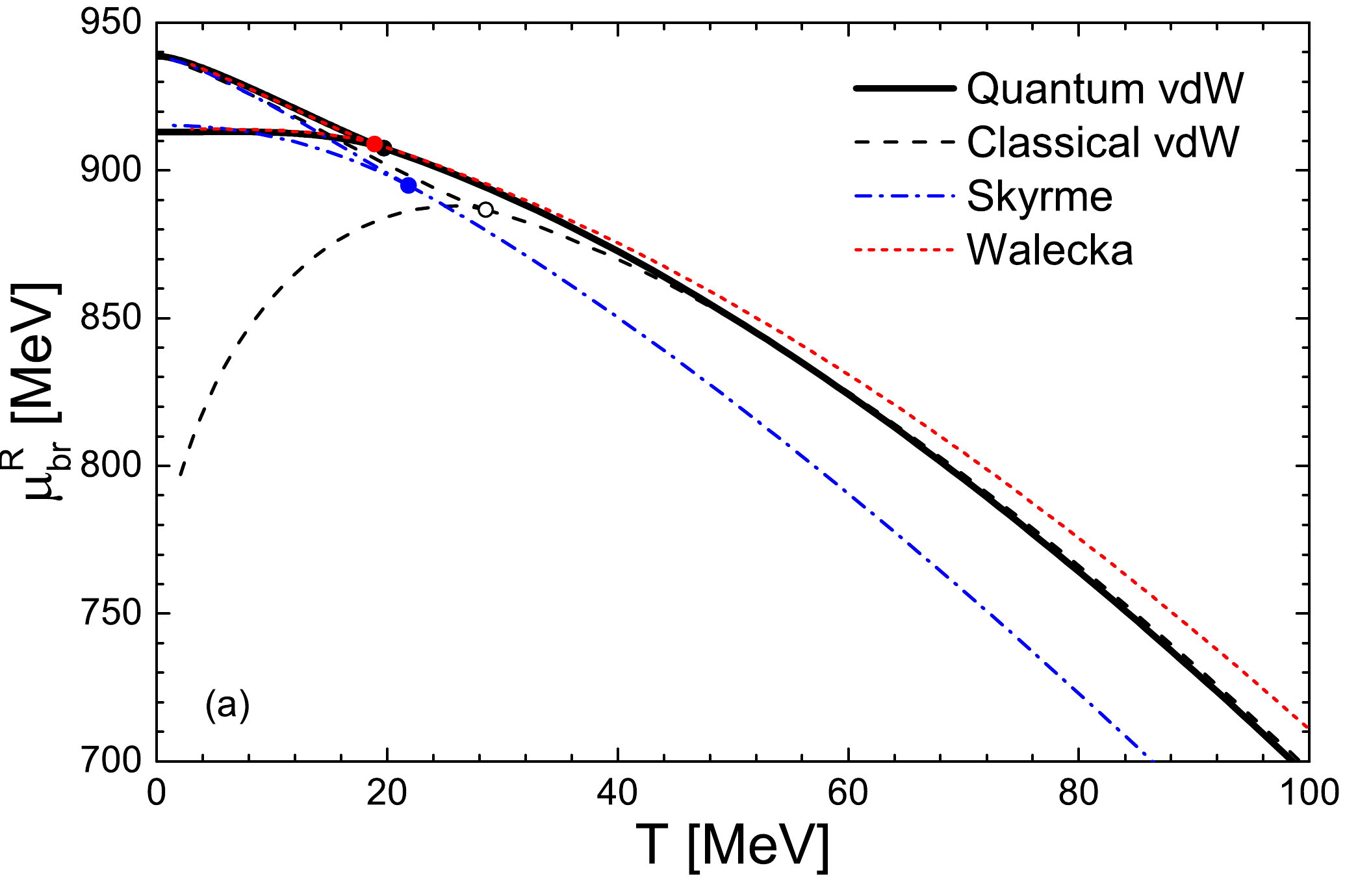}
  \includegraphics[width=.49\textwidth]{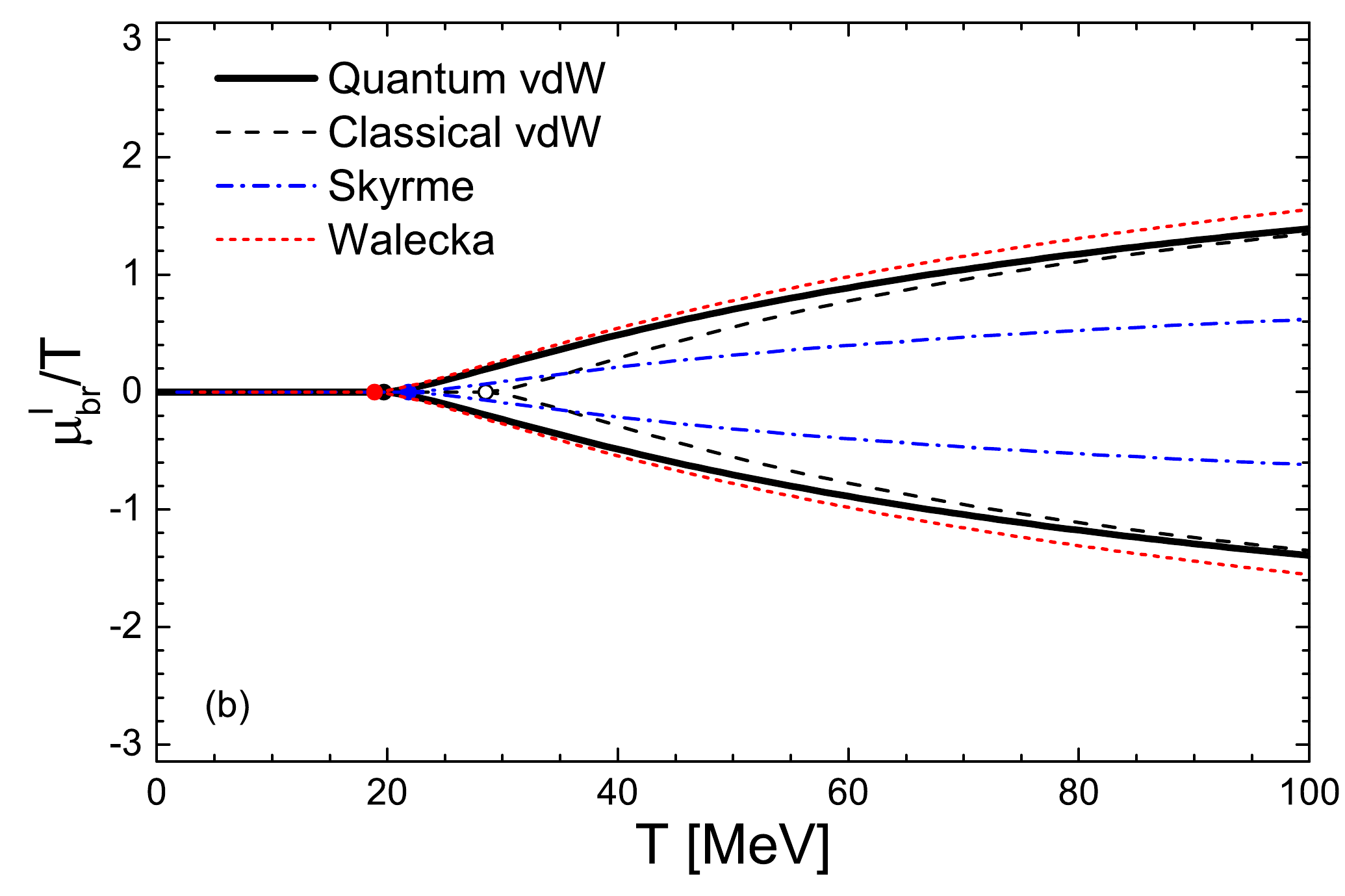}
  \caption{
   Temperature dependence of (a) $\mu_{\rm br}^{R} \equiv \text{Re} [ \mu_{\rm br}]$ and (b) $\mu_{\rm br}^{I}/T \equiv \text{Im} [ \mu_{\rm br} / T]$ evaluated for a system of interacting nucleons through the classical van der Waals equation~(dashed black lines), the quantum van der Waals equation~(solid black lines),
   the Skyrme mean-field model~(dot-dashed blue lines)
   and the Walecka relativistic mean-field model~(dotted red lines).
   The points depict the respective locations of the nuclear matter critical points in the corresponding models.
  }
  \label{fig:NMcompare}
\end{figure*}

A comparison between the classical and quantum vdW models can clarify the role of the Fermi statistics.
The comparison is exhibited in Fig.~\ref{fig:NMcompare}, where the temperature dependencies of real and imaginary parts of $\mu_{\rm br}$ are depicted.
Both models exhibit qualitatively similar behavior. 
The classical vdW model does not yield an accurate description of $\mu_{\rm br}$ at small temperatures. This is an expected artefact of neglecting the quantum statistics.
Classical vdW model results approach the QvdW model at large temperatures, where effects of quantum statistics become negligible.

\subsection{Skyrme model}

To cross-check the robustness of results obtained in the framework of (quantum) vdW model we consider thermodynamic branch points in 
two
alternative models of nuclear matter.
In the Skyrme model of nuclear matter the attractive and repulsive interactions are modeled through a mean-field \cite{Bender:2003jk,Stone:2006fn},
\eq{\label{eq:MFSk}
u_{\rm sk}(n) = - \alpha \left(\frac{n}{n_0}\right) + \beta \, \left(\frac{n}{n_0}\right)^{\gamma},
}
which shifts the single-particle energy levels.
Here the first term corresponds to intermediate-range attractive interactions and the second term to short-range repulsive interactions.
The nucleon number density is given by a self-consistent equation,
\eq{\label{eq:nSk}
n(T, \mu) = n_{\rm id}[T, \mu - u_{\rm sk}(n)]~.
}

Here we use the following parameter values: $n_0 = 0.16$~fm$^{-3}$, $\gamma = 2$, $\alpha \simeq 122.6$~MeV and $\beta \simeq 70.4$~MeV.
These parameter values yield the binding energy of 16~MeV in the nuclear ground state at $n = n_0 = 0.16$~fm$^{-3}$.
The $\gamma = 2$ value corresponds to the so-called hard Skyrme equation of state, with an incompressibility modulus of $K_0 \simeq 380$~MeV.
The CP is located in this model at $T_c = 21.9$~MeV, $n_c = 0.06$~fm$^{-3}$~($\mu_c \simeq 895$~MeV).

The derivative $(\partial \mu / \partial n)_T$ can be evaluated from Eq.~\eqref{eq:nSk} in a fairly straightforward manner.
The branch point equation $(\partial \mu / \partial n)_T = 0$ reads
\eq{\label{eq:brSk}
\left[ \alpha \left(\frac{n_{\rm br}}{n_0}\right) - \beta \, \gamma \, \left(\frac{n_{\rm br}}{n_0}\right)^{\gamma} \right] \frac{\omega_{\rm id}[T, \mu - u_{\rm sk}(n_{\rm br})]}{T} = 1.
}
In practice, Eq.~\eqref{eq:brSk} is solved numerically for a quantity $\mu^*_{\rm br} \equiv \mu - u_{\rm sk}(n_{\rm br})$, as Eq.~\eqref{eq:nSk} gives $n_{\rm br}$ as an explicit function of $\mu^*_{\rm br}$.
As for the QvdW model, the CP location is used as a starting point of the numerical procedure to determine the temperature dependence of the branch points.

\subsection{Walecka model}

The last nuclear matter model under consideration is the Walecka model~\cite{Walecka:1974qa,Serot:1984ey}, which is one of the simplest examples of a relativistic mean field theory.
The attractive and repulsive interactions are modeled through exchange of scalar $\sigma$ and vector $\omega$ mesons, respectively. The mesonic fields are treated in a mean-field approximation.
The interactions lead to an effective shift of the chemical potential $\mu \to \mu^*$ and mass $m \to m^*$ of nucleons, leading to the following form of the grand-canonical thermodynamic potential~(pressure)\footnote{Here we neglect the contribution of anti-nucleons which is 
small
in the nuclear matter region of the phase diagram.}
\eq{\label{eq:pWal}
p(T,\mu) = p_{\rm id}(T,\mu^*;m^*) + \frac{(\mu-\mu^*)^2}{2 c_v^2} - \frac{(m-m^*)^2}{2 \,c_s^2}~.
}
Here $c_s^2 > 0$ and $c_v^2 > 0$ are the coupling parameters corresponding to attractive and repulsive interactions, respectively.
The effective chemical potential $\mu^*$ and effective mass $m^*$ are determined from gap equations:
\eq{\label{eq:GapWal}
\left( \frac{\partial p}{\partial \mu^*} \right)_{m^*} = 0 & \quad \Longleftrightarrow \quad & \mu - \mu^* = c_v^2 \, n_{\rm id}(T,\mu^*;m^*), \\
\label{eq:GapWal2}
\left( \frac{\partial p}{\partial m^*} \right)_{\mu^*} = 0 & \quad \Longleftrightarrow \quad & m - m^* = c_s^2 \, n^s_{\rm id}(T,\mu^*;m^*).
}
Here $n^s_{\rm id}$ is the scalar density of an ideal Fermi gas of nucleons.
The particle number density is
\eq{\label{eq:nWal}
n(T,\mu) = n_{\rm id}(T, \mu^*; m^*).
}

The values of coupling parameters are determined from the nuclear ground state properties~(see Ref.~\cite{Poberezhnyuk:2017yhx} for details): $c_s^2 = 14.6$~fm$^2$ and $c_v^2 = 11.0$~fm$^2$.
The model predicts nuclear matter CP at $T_c = 18.9$~MeV, $n_c = 0.07$~fm$^{-3}$~($\mu_c \simeq 909$~MeV).

The branch points are determined through Eq.~\eqref{eq:br}.
In order to evaluate $(\partial \mu / \partial n)_T$ we first note that
$\mu = \mu^* + c_v^2 n$, as follows from Eqs.~\eqref{eq:GapWal} and \eqref{eq:nWal}.
Therefore,
\eq{
(\partial \mu / \partial n)_T = (\partial \mu^* / \partial n)_T + c_v^2.
}

$(\partial \mu^* / \partial n)_T$ is determined by applying the $(\partial/\partial n)_T$ derivative to the gap equations~\eqref{eq:GapWal} and \eqref{eq:GapWal2}, and solving the resulting system of linear equations for $(\partial \mu^* / \partial n)_T$ and $(\partial m^* / \partial n)_T$:
\eq{
(\partial \mu^* / \partial n)_T = \frac{1 + c_s^2 \, \partial_{m^*} n_s^*}{\partial_{\mu^*} n^* + c_s^2 \, ( \partial_{\mu^*} n^* \, \partial_{m^*} n_s^* - \partial_{\mu^*} n_s^* \, \partial_{m^*} n^* )}.
}
Here $n^* \equiv n_{\rm id}(T,\mu^*;m^*)$ and $n_s^* \equiv n^s_{\rm id}(T,\mu^*;m^*)$.

The branch points equation~\eqref{eq:br} reads
\eq{
1 & + c_s^2 \, \partial_{m^*} n_s^* + c_v^2 \, \partial_{\mu^*} n^* \nonumber \\
& \quad = c_s^2 \, c_v^2 \, (\partial_{\mu^*} n_s^* \, \partial_{m^*} n^* - \partial_{\mu^*} n^* \, \partial_{m^*} n_s^* ).
}
This equation is solved numerically to determine $\mu^*_{\rm br}$~(the gap equation~\eqref{eq:GapWal2} is used to relate $m^*$ and $\mu^*$).

\subsection{Comparison between models}

Figure~\ref{fig:NMcompare} depicts the temperature dependence of the real and imaginary parts of the branch point chemical potential evaluated in the four considered models of nuclear matter.
The qualitative behavior in all models is
consistent with the analytic expectations obtained within the classical vdW model
in Sec.~\ref{sec:vdwcl}.
As mentioned before, at small temperatures the $\mu_{\rm br}^R$ values in the classical vdW model are quite different from all other models. This is an artifact due to the absence of Fermi statistics in the classical vdW model.

The large temperature behavior is qualitatively similar in all models considered.
Quantitatively, the QvdW and Walecka models are very similar while the Skyrme model yields a stronger temperature dependence of the real part $\mu_{\rm br}^R$ and a milder temperature rise of the imaginary part $\mu_{\rm br}^{I}/T$.

At large temperatures the classical and quantum vdW models give almost identical results.
Effects of Fermi statistics are negligible at $T \gtrsim 100$~MeV.
The analytic results of Sec.~\ref{sec:vdwcl} describe $\mu_{\rm br}^R$ and $\mu_{\rm br}^I$ very accurately in this regime.

\begin{figure*}[t]
\centering
\includegraphics[width=.49\textwidth]{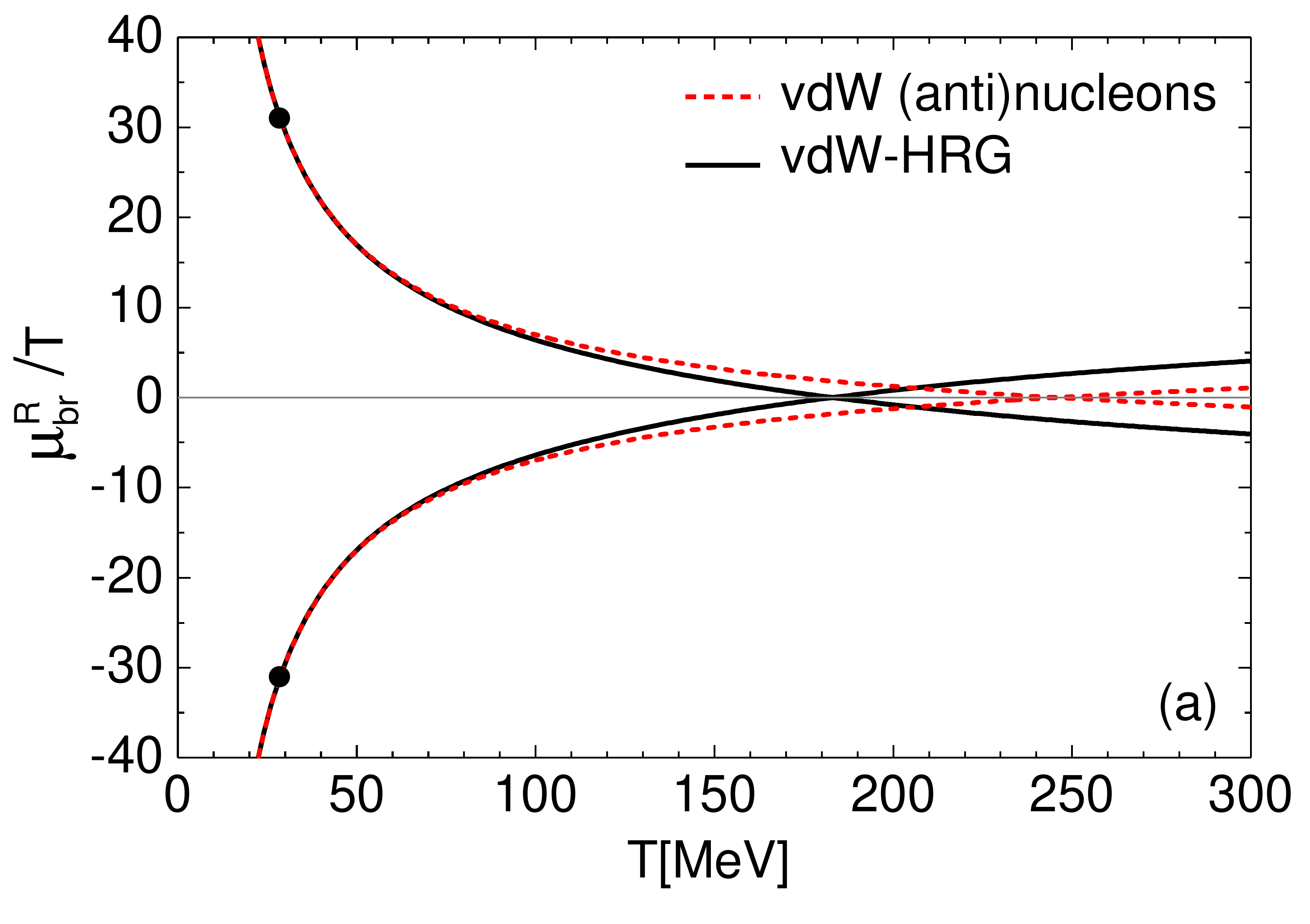}
\includegraphics[width=.49\textwidth]{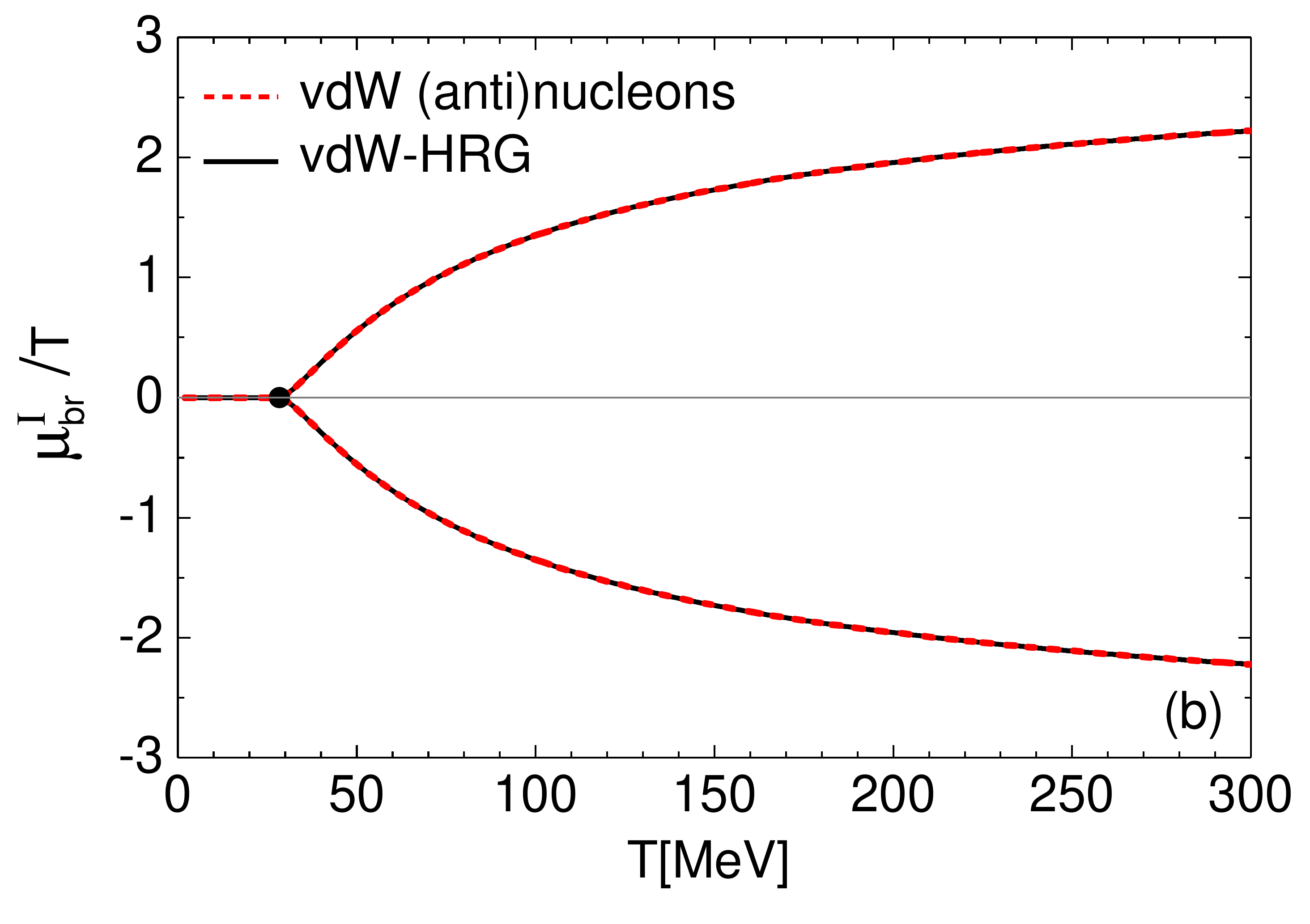}
\caption{
The temperature dependence of the (a) real part of the limiting branch cut singularity, and (b) imaginary part of the limiting branch cut singularity.
Calculations are performed for the vdW-HRG model~(solid black lines) and for the vdW model with (anti)nucleons only~(dashed red lines).
Maxwell-Boltzmann statistics is considered in both cases.
The circles correspond to the critical points of, respectively, nuclear matter 
and nuclear anti-matter.
}
\label{fig:mubrvdwHRG}
\end{figure*}

\section{Hadron resonance gas and nuclear matter}
\label{sec:vdWHRG}

\subsection{Branch points}

At temperatures $T \gtrsim 100$~MeV, which are probed by relativistic heavy-ion collisions and studied in finite-temperature lattice gauge theory, excitations of hadronic degrees of freedom other than nucleons cannot be neglected.
The hot hadronic phase is typically modeled in the framework of the hadron resonance gas~(HRG) model.
The standard HRG model does not usually incorporate nuclear matter properties and the associated liquid-gas criticality.

Here we employ a vdW-HRG model, which has been introduced in Ref.~\cite{Vovchenko:2016rkn} as a ``minimum'' extension of the HRG model to incorporate the nuclear liquid-gas phase transition into a HRG picture. In the present paper we follow this ``minimum'' extension.
The vdW-HRG model incorporates vdW interactions for all baryon-baryon~(and, by symmetry, all antibaryon-antibaryon) pairs.
The parameters $a$ and $b$ are taken to be the same for all baryon pairs. 
Their values are equal to the vdW parameters of nucleons~(Sec.~\ref{sec:num}).
This ensures that the vdW-HRG model reduces to the vdW model of nuclear matter in Sec.~\ref{sec:num} when the contributions of baryonic resonances become negligible, as is the case for the low $T$, large $\mu_B$ nuclear matter region of the phase diagram.
 
We do not include vdW terms for baryon-antibaryon pairs as baryon-antibaryon interactions at short range are dominated by annihilations rather than by a repulsive core as in baryon-baryon interactions.
Finally, most of the known meson-meson and meson-baryon scatterings are dominated by resonance formation. Such interactions are already incorporated in a HRG picture by including resonances as separate particles. 
Note also that our particle list has no resonances with $|B| = 2$, therefore, there is no double-counting of attractive interactions between baryon-baryon and antibaryon-antibaryon pairs.

% Note that the vdW interactions is only one of the possible ways to implement the nuclear criticality within the HRG model (see Sec.~\ref{sec:num}).
% However, different models where considered in Sec.~\ref{sec:num} to show moderate model dependence of the behavior of the thermodynamic branch points, at least on the mean field level. 
% This allows us to focus on one of the implementations, namely the vdW interactions, in the following. 

The pressure in the vdW-HRG model reads:
\eq{\label{p-HRG}
p(T,\mu) = p_M(T,\mu) + p_B(T,\mu) + p_{\bar{B}}(T,\mu),
}
with
\begin{align}
p_M(T, \mu) & =
\sum_{j \in M} p_{j}^{\rm id} (T, \mu_j) \\
\label{eq:PB}
p_B(T, \mu) & =
\sum_{j \in B} p_{j}^{\rm id} (T, \mu_j^{B*}) - a\,n_B^2 \\
\label{eq:PBBar}
p_{\bar{B}}(T,\mu) & =
\sum_{j \in \bar{B}} p_{j}^{\rm id} (T, \mu_j^{\bar{B}*}) - a\,n_{\bar{B}}^2,
\end{align}
where $M$ stands for mesons, $B(\bar{B})$ for (anti)baryons, 
$\mu=(\mu_B,\mu_S,\mu_Q)$ are the chemical potentials 
for net baryon number $B$, strangeness $S$, and electric charge $Q$,
$\mu_j^{B(\bar{B})*} = \mu_j - b\,p_{B(\bar{B})} - a\,b\,n_{B(\bar{B})}^2 + 2\,a\,n_{B(\bar{B})}$ where
$\mu_j = B_j \, \mu_B + S_j \, \mu_S + Q_j \, \mu_Q$ is the chemical potential for baryon species $j$, with $B_j$, $S_j$, and $Q_j$ being its corresponding quantum numbers.
$n_B$ and $n_{\bar{B}}$ are total densities of baryons and antibaryons, respectively.

We neglect the quantum statistical effects for 
baryons and anti-baryons
in the following.
As was shown in Sec.~\ref{sec:num} this is a good approximation for temperatures $T \gtrsim 80$~MeV (see Fig.~\ref{fig:NMcompare}).
For $\mu_Q = \mu_S = 0$, the (anti)baryon densities $n_{B(\bar{B})}(T, \mu)$ are defined by the transcendental equation:
\eq{\label{eq:density}
b  \, \phi_B(T) e^{\pm \mu_B/T} = \frac{b\nBB\exp\left[ \frac{b  \nBB}{1 - b  \nBB} - \frac{2  a  \nBB}{T} \right]}{1 - b \nBB}  .
}
Here
\eq{\label{eq:phiB}
\phi_B(T) = \sum_{i \in B} \frac{d_i \, m_i^2 \, T}{2 \pi^2} \, K_2(m_i/T)~.
}
The sum in Eq.~\eqref{eq:phiB} runs over all baryons in the HRG.

Densities $\nBB$ are multivalued functions of $\mu_B$.
Both baryons and antibaryons lead to an appearance of branch points.
Due to the charge conjugation parity symmetry, the corresponding branch points are related to each other through a transformation $\mu_B \to - \mu_B$. 
The branch points of $\nBB$ are defined as
\eq{\label{eq:BranchingPoints}
\left. \frac{d \mu_B}{d \nBB} \right|_{\mu_B = \mu_B^{\rm br}} = 0.
}

The branch point coordinates are determined through the relations for the classical vdW equation~(Sec.~\ref{sec:vdwcl}) with a substitution $\phi (T) \to \phi_{B(\bar{B})}(T)$.
Figure~\ref{fig:mubrvdwHRG} depicts the temperature dependence of the real and imaginary parts of the branch cut singularities associated with the nuclear liquid-gas transition, evaluated within the vdW-HRG model~(solid black lines) and the vdW model with (anti)nucleons only~(dashed red lines). Only limiting, i.e., closest to the $\mu_B=0$ expansion point, singularities are presented. Two symmetric lines in ($a$) correspond to baryons and antibaryons. Imaginary parts of two complex-conjugated singularities, presented in ($b$), are equal for baryons and antibaryons.
Circles represent the critical points of baryonic and antibaryonic matter.
It is seen that the real part decreases with temperature and crosses zero at about $T \simeq 180$~MeV. 
This implies that vdW interactions become relevant even close to $\mu_B = 0$ at sufficiently large temperatures. This indeed was demonstrated for a number of thermodynamic quantities in Ref.~\cite{Vovchenko:2016rkn}.
The addition of the baryonic resonances leads to a faster decrease of $\mu_{\rm br}^R/T$ towards zero. 
On the other hand, the resonances do not affect the behavior of the imaginary part $\mu_{\rm br}^I/T$, at least not within the vdW-HRG model used.

\begin{figure*}
  \includegraphics[width=.49\textwidth]{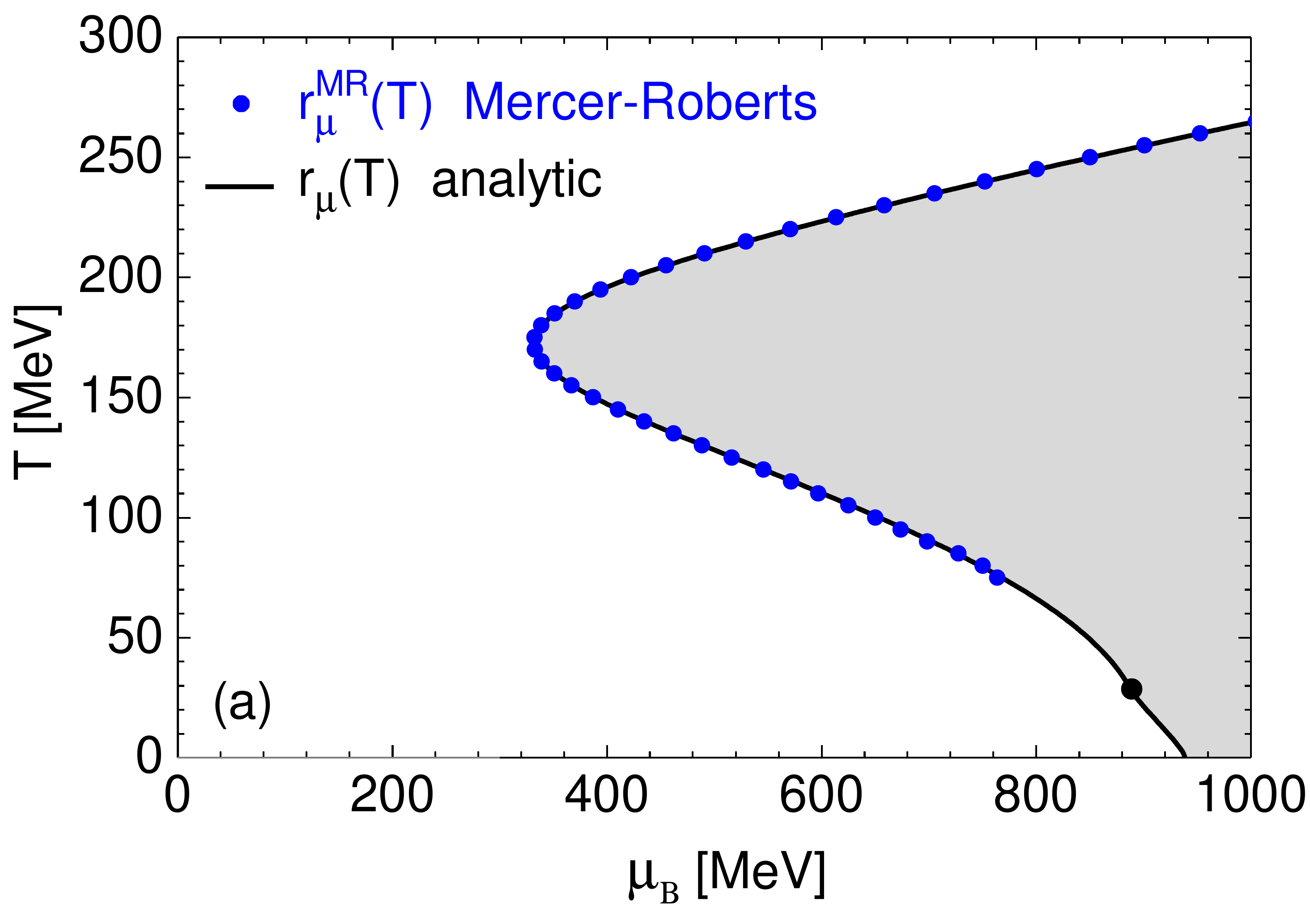}
  \includegraphics[width=.49\textwidth]{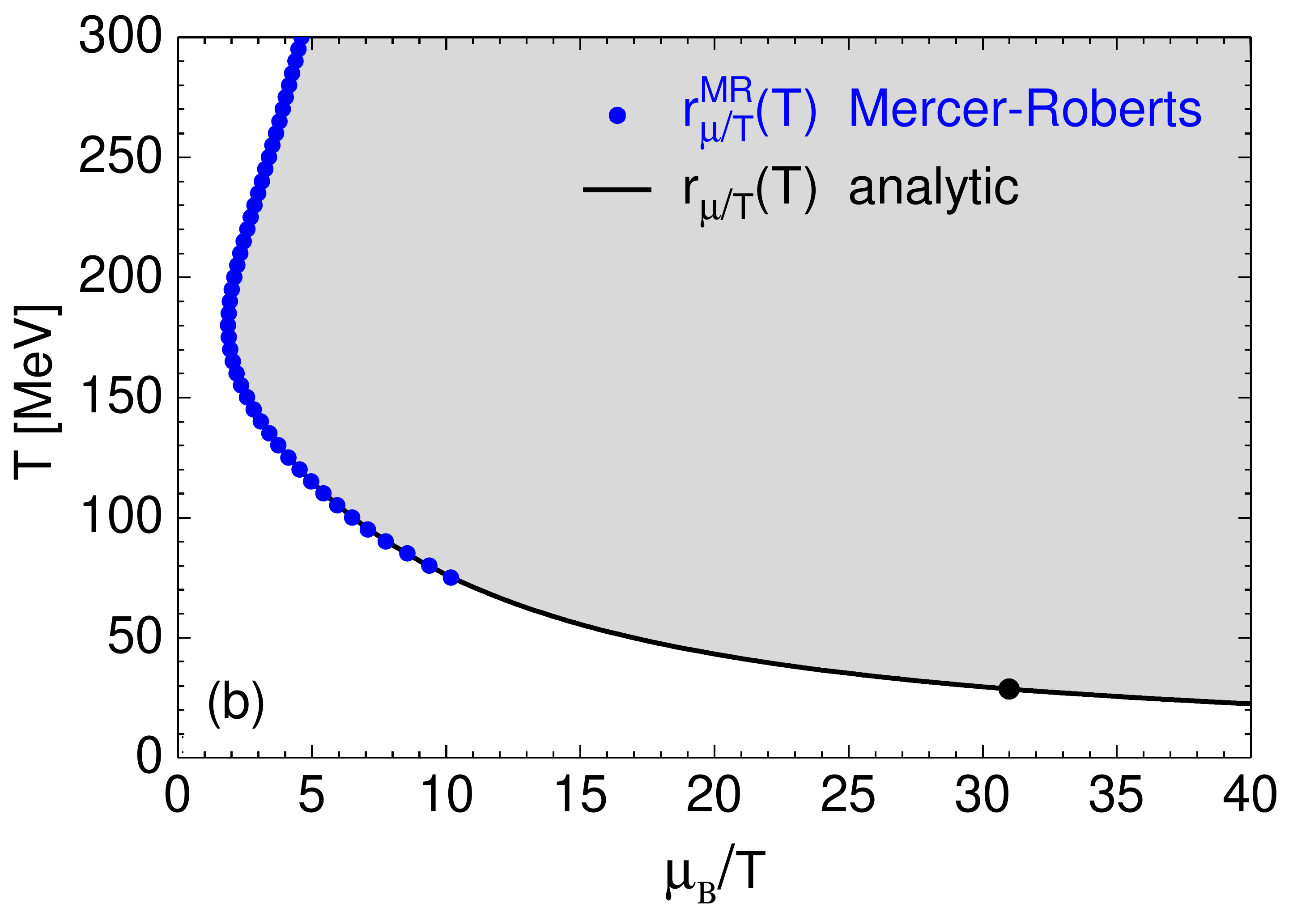}
  \caption{
   The 
   temperature dependence of  ($a$) the radius of convergence $r_\mu$ and ($b$) the radius of convergence $r_{\mu/T}$ calculated in vdW-HRG model (solid line) numerically using analytic formulae Eqs.~(\ref{eq:BranchingPoints}) and (\ref{rmu-modul}); and (blue dots) using the Mercer-Roberts radius of convergence estimator (\ref{eq:MR}) for Taylor expansion (\ref{eq:TaylorExp}).
Shaded areas represent that region of the phase diagram where the Taylor expansion (\ref{eq:TaylorExp}) of the pressure function does not converge.
   The black circle corresponds to the critical point of nuclear matter in the classical vdW model.
  }
  \label{fig:rmuMR}
\end{figure*}

\subsection{Taylor expansion}

The presence of thermodynamic branch points leads to a number of consequences regarding the analytic properties of QCD. Of particular interest is the Taylor expansion of the QCD pressure:
\eq{\label{eq:TaylorExp}
\frac{p(T,\mu_B) - p(T,0)}{T^4} = \sum_{n=1}^{\infty} \, \frac{\chi_{2n}^B (T)}{(2n)!} \, \left( \frac{\mu_B}{T}\right)^{2n}~.
}
Here $\chi_{2k}^B (T) = \partial^{2k} (p/T^4) / \partial (\mu_B/T)^{2k}|_{\mu_B = 0}$ are the baryon number susceptibilities evaluated at $\mu_B = 0$. 
Expansion includes only even orders of chemical potential as follows from the charge conjugation  parity symmetry of QCD. 

The series~(\ref{eq:TaylorExp}) converges inside a circle in the complex $\mu_B/T$ plane.
The convergence is limited by a singularity closest to the expansion point, which lies on the border of the circle.
A CP is an example of such singularity. 
A particular feature of a CP is that the singularity lies on the real axis, implying that Taylor expansion coefficients are asymptotically positive at the critical temperature.
This fact is used in various attempts to constrain the location of the QCD CP using lattice QCD, by evaluating a number of leading order Taylor expansion coefficients at $\mu_B = 0$, verifying that all available coefficients are positive, and using various radius of convergence estimators~\cite{DElia:2016jqh,Datta:2016ukp,Bazavov:2017dus}.
Note that a divergent Taylor expansion can appear even without the presence of physical phase transitions, e.g. in systems with repulsive interactions only~\cite{Taradiy:2019taz}.

The thermodynamic singularities associated with the nuclear liquid-gas transition do limit the convergence range of Taylor expansion in the vdW-HRG model.
The expected radius of convergence in the vdW-HRG model is given by
\eq{\label{rmu-modul}
r_{\mu} = |\mu_B^{\rm br}| = \sqrt{ [ \text{Re} (\mu_B^{\rm br}) ]^2 + [\text{Im} (\mu_B^{\rm br}) ]^2 }.
}
Here $\mu_B^{\rm br}$ is the location of the limiting singularity.
At $T > T_c$ this corresponds to the crossover singularities~[Eq.~\eqref{eq:firstroot}], which both lie at the same distance from $\mu_B = 0$.
At $T = T_c$ this is the nuclear matter CP.
At $T < T_c$ the limiting singularity is the spinodal point which separates the gaseous and mechanically unstable nuclear phases~(the right solid curve in Fig.~\ref{fig:NMQvdW}).\footnote{The branch point at the boundary of the liquid and the mechanically unstable phase~(the left solid curve in Fig.~\ref{fig:NMQvdW}) does not limit the radius of convergence of Taylor expansion, despite its smaller $\mu_B^{\rm br}$ value.
The reason is that this branch point lies on a Riemann surface different from the one where the expansion point $\mu_B = 0$ is.
The radius of convergence is unaffected by the third root~[Eq.~\eqref{eq:nbr3}] for the same reason.
}

Figure~\ref{fig:rmuMR} depicts the temperature dependence of the radius of convergence in $\mu_B$ and in dimensionless $\mu_B/T$ variables. 
For crossover temperatures, $T \sim 140-170$~MeV, the radius of convergence becomes as small as $r_{\mu/T} \sim 2-3$.
This indicates a real possibility that convergence properties of Taylor expansion in full QCD at crossover temperatures might be determined by the remnants of the nuclear liquid-gas transition, which manifest themselves in a form of singularities in the complex plane.

We cross-check our vdW-HRG model results by analyzing the convergence properties of Taylor expansion in this model directly.
First, we analyze the convergence radius from the behavior of net baryon susceptibilities $\chi_k^B$ at $\mu_B = 0$.
We calculate $\chi_k^B$ up to the order $\chi_{120}^B$
% ~\vv{Oleg, put the correct number here.} 
numerically, using an efficient algorithm described in the Appendix.
The computed $\chi_k^B$ values are then used to determine $r_{\mu/T}$ through various estimators.

The so-called ratio estimator, $r^{\rm RE}_n = \left| c_n / c_{n+1}\right|^{1/2}$ with $c_n \equiv \chi_{2n} / (2n)!$, fails to provide a useful estimate of $r_{\mu/T}$.
This is a consequence of the fact that limiting singularity lies in the complex plane, with a non-zero imaginary part $\mu_{\rm br}^I/T \neq 0$. In such a case the ratio estimator does not converge~\cite{Vovchenko:2017gkg,Giordano:2019slo}.

An accurate estimate for the radius of convergence is obtained using the Mercer-Roberts estimator~\cite{MercerRoberts},
\eq{\label{eq:MR}
r^{\rm MR}_n = \left| \frac{c_{n+1} \, c_{n-1} - c_n^2 } { c_{n+2} \, c_n - c_{n+1}^2 }\right|^{1/4},
}
and the so-called Domb-Sykes presentation~\cite{DombSykes,DombSykes2}~(see details in Ref.~\cite{Vovchenko:2017gkg}). 
The resulting values of $r_\mu$ and $r_{\mu/T}$ using 120 Taylor expansion coefficients are depicted by blue symbols in Fig.~\ref{fig:rmuMR}.
These values agree with the prior analytic expectations shown by the solid lines,
indicating that the Mercer-Roberts estimator converges to the correct value.
We also analyze how many coefficients are needed to obtain a meaningful estimate of $r_{\mu/T}$.  
The calculations suggest that $r_{\mu/T}$ at $T=100-200$~MeV can be estimated with a 10\% accuracy using $5-10$ nonzero Taylor expansion coefficients (see Tab.~\ref{tab:rn}). 
We note that performance of estimating $r_{\mu/T}$ can be improved by considering a modified Mercer-Roberts estimator of Ref.~\cite{Giordano:2019slo}:
\eq{
r^{\rm MMR}_n = \left| \frac{(n+1)(n-1)c_{n+1} \, c_{n-1} - n^2 c_n^2 } { (n+2)n c_{n+2} \, c_n - (n+1)^2 c_{n+1}^2 }\right|^{1/4}.
}
This is shown in the 2nd row of Tab.~\ref{tab:rn}.

\begin{center}
\begin{table}
\begin{tabular}
{| l | c | c | c | c |}
\hline
\hline
 Estimator&                          ~100~MeV  &~150~MeV~&~170~MeV~&~200~MeV~\\
\hline
        MR       & $\chi_{16}^B$  & $\chi_{16}^B$   & $\chi_{10}^B$ & $\chi_{20}^B$ \\
MMR         & $\chi_{22}^B$ & $\chi_{10}^B$ & $\chi_{8}^B$ & $\chi_{14}^B$ \\
  \hline
   \hline
\end{tabular}
\caption{
\label{tab:rn} 
Depicts the number of leading Taylor expansions coefficients
needed to extract the radius of convergence $r_\mu$ within $10\%$ of the true value through the 
Mercer-Roberts (first row) and modified Mercer-Roberts~(second row) estimators in the vdW-HRG model.
}
\end{table}
\end{center}

\begin{figure}[t]
  \centering
  \includegraphics[width=.49\textwidth]{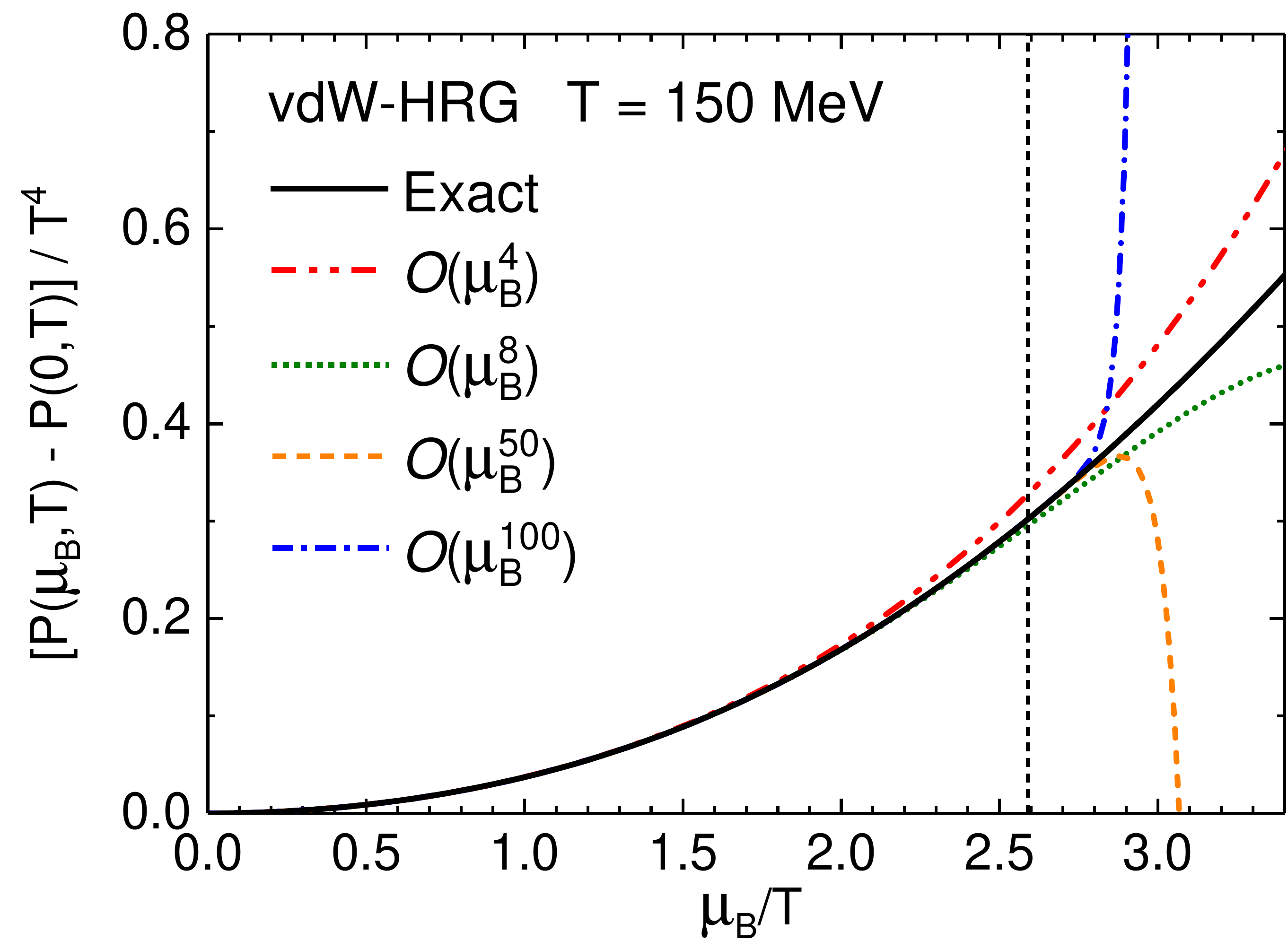}
  \caption{The dependence of the subtracted scaled pressure $[p(T,\mu_B)-p(T,0)]/T^4$ on $\mu_B/T$, as calculated within the vdW-HRG model at $T = 150$~MeV using the numerical solution~[Eq.~\eqref{eq:density}]~(solid black line) and the Taylor expansion 
truncated at $\chi_{4}^B$~(dash-dotted red line), $\chi_{8}^B$~(dotted green line), $\chi_{50}^B$~(dashed yellow line) and $\chi_{100}^B$~(dash-dotted blue line).
  The vertical dashed line corresponds to the value of the convergence radius $r_{\mu/T} \simeq 2.6$.}
  \label{fig:TaylorTest}
\end{figure}

Next, we study the convergence properties of the Taylor expansion by comparing the pressure isotherms evaluated using a truncated Taylor expansion around $\mu_B = 0$ and through a full numerical calculation.
Figure~\ref{fig:TaylorTest} shows this comparison for the $T = 150$~MeV isotherm, where a subtracted scaled pressure $[p(T,\mu_B)-p(T,0)]/T^4$ as a function of $\mu_B/T$ is analyzed.
The truncated Taylor expansion describes well the full result for $\mu_B/T < r_{\mu/T}$ as well as in a small region beyond $r_{\mu/T}$.
We verified that this small region shrinks towards zero as more and more expansion terms are included, and that divergence of the series at $\mu_B/T > r_{\mu/T}$ becomes more and more evident.
Thus, for $\mu_B/T > r_{\mu/T}$ the Taylor expansion can at best be only viewed as an asymptotic series.

%The presented calculation illustrates how the presence of the nuclear liquid-gas transition can affect the convergence properties of Taylor expansion.
The present calculation incorporates only hadronic degrees of freedom.
Of course, mechanisms other than the nuclear liquid-gas transition are present in full QCD, 
which affect the analytic properties of the thermodynamic potential and which are not covered within the vdW-HRG model.
This includes, for instance, the QCD transition to quark-gluon degrees of freedom and the associated chiral criticality.
Another known QCD transition is the Roberge-Weiss transition at imaginary chemical potential~\cite{Roberge:1986mm}, occurring at temperatures $T > T_{\rm RW}$, where $T_{\rm RW} \sim 208$~MeV~\cite{Bonati:2016pwz}.
These mechanisms will yield additional restrictions on the radius of convergence of Taylor expansion, in addition to those due to nuclear liquid-gas transition alone that we study here.
The work reported here, on the other hand, merely demonstrates how the presence of the nuclear liquid-gas transition alone can affect the convergence properties of the Taylor expansion.
For similar reasons we do not perform here a comparison of the higher-order cumulants $\chi_{2k}^B$ with the lattice data, but merely use these quantities to analyze the behavior of the Taylor-expanded pressure.

It should also be noted that the liquid-gas transition has been treated in this work on a mean-field level, and the associated critical behavior corresponds to the mean-field universality class.
Going beyond the mean-field approximation can modify the universality class and the nature of the singularity associated with a critical point of a phase transition~\cite{Stephanov:2006dn}, which, in turn, will modify the behavior of the higher-order susceptibilities $\chi_{2k}^B$.
Studying the thermodynamic singularities beyond the mean field level can, therefore, be an interesting future endeavor, which can be achieved e.g. using renormalization group methods.

%chiral and deconfinement aspects of the QCD transition to quark-gluon degrees of freedom that we do not consider here. 

% Note that although the vdW-HRG model is sufficient to illustrate that the nuclear liquid-gas transition affects the convergence properties of the Taylor expansion,
% for a quantitative description of the lattice data on the higher-order cumulants at high temperatures, $T>150$~MeV, the model needs to incorporate the  quark-gluon degrees of freedom, which become more and more relevant as the temperature is increased. 

\section{Summary}
\label{sec:summary}

The presence of the nuclear liquid-gas transition at temperatures $T \lesssim 20$~MeV in QCD leads to the emergence of thermodynamic branch points in the QCD grand potential.
These branch points correspond to the spinodals of the first-order phase transition at $T < T_c$, to the critical point at $T = T_c$, and to crossover singularities in the complex $\mu_B$ plane at $T > T_c$.
This qualitative result, obtained analytically within the classical vdW equation, is generic for any arbitrary mean-field  description of nuclear matter, as follows from the universality argument for critical behavior.
From a quantitative point of view, this behavior of the branch points exhibits mild model dependence at moderate temperatures $T \lesssim 100$~MeV, whereas the inclusion of all other hadronic degrees of freedom proves also important at higher temperatures, $T \sim 130-180$~MeV.

The van der Waals hadron resonance gas model analysis implies that signals from the nuclear liquid-gas transition are clearly visible in analytic properties of QCD even at crossover temperatures and moderate baryochemical potentials.
In particular, the radius of convergence of a Taylor expansion reaches in the vdW-HRG model values as small as $r_{\mu/T} \sim 2-3$ for temperatures $T \sim 140-170$~MeV. 
Such high temperatures are typically assumed to only be associated with the chiral crossover transition at $\mu_B = 0$.
However, the present results show that the radius of convergence of the Taylor expansion in QCD at these temperature exhibits clearly the remnants of the nuclear liquid-gas transition, at a region where we expected the signals of chiral criticality. 
If the hypothetical chiral critical point is located deeply in baryon-rich matter, 
as indicated by a recent analysis of QCD thermodynamics within the chiral mean-field approach~\cite{Motornenko:2019arp},
the attempts to locate the QCD CP by using the Taylor expansion method must take great care to distinguish the supposed signals of the conjectured chiral CP from the well established nuclear matter liquid-vapor CP.

\section*{Acknowledgments}
The work of M.I.G. is supported 
by the Program of Fundamental Research of the Department of Physics and Astronomy of the National Academy of Sciences of Ukraine.
H.St. acknowledges the support through the Judah M. Eisenberg Laureatus Chair by Goethe University  and the Walter Greiner Gesellschaft, Frankfurt.

%\end{spacing}

\begin{widetext}

\section*{Appendix}
\subsection{On the evaluation of baryon number susceptibilities in the vdW-HRG model}
 An even order baryon number susceptibility, $\chi_{2m}$, is expressed at $\mu_B=0$  through the $(2m-1)$ order derivative of the baryonic density, $n_B$, w.r.t. $\mu_B/T$:
\eq{\label{chi2m-n}
    \chi_{2m} (T) = 
   \left. \frac{\partial^{2m} (p/T^4)}{\partial (\mu_B/T)^{2m}}\right\vert_{\mu_B=0}=
    \left. \frac{1}{T^3}\frac{\partial^{2m-1}(n_{B}-n_{\bar{B}})}{\partial\left(\mu_B/T\right)^{2m-1}}\right\vert_{\mu_B=0}=
   \left. \frac{2}{T^3}\frac{\partial^{2m-1}n_{B}}{\partial\left(\mu_B/T\right)^{2m-1}}\right\vert_{\mu_B=0}.
}
Here we used Eq.~(\ref{p-HRG}), the fact that 
$\partial p_{B(\bar{B})}/\partial \mu_B=\pm n_{B}$ and 
$\partial^k n_{\bar{B}}/\partial (\mu_B)^k|_{\mu_B=0}=(-1)^{k+1} \partial^k n_{B}/\partial (\mu_B)^k|_{\mu_B=0}$.

To find an arbitrary order derivative $\partial^k n_B/\partial(\mu_B/T)^k|_{\mu_B=0}$ we rewrite Eq.~(\ref{eq:density}) for baryon density in the following form: 
\eq{\label{f=g}
f(T, \mu_B)=g(n_B),
}
 where in the vdW-HRG model
\eq{\label{eq:f-g}
f(T, \mu_B)={\rm ln}[b  \, \phi_B(T)]+\mu_B/T,~~~~~~~~ {\rm and}~~~~~~~~ g(n_B)={\rm ln}[b n_B]-{\rm ln}[1-b n_B]+\frac{b  n_B}{1 - b  n_B} - \frac{2  a  n_B}{T},
}
are the logarithms of, respectively, l.h.s and r.h.s. of Eq.~(\ref{eq:density}). Applying the Faà di Bruno's formula 
to Eq.~(\ref{f=g}) gives,
\eq{\label{faa}
  \frac{\partial^n f}{\partial(\mu_B/T)^n}=\sum_{k=1}^{n}\frac{\partial^kg}{\partial (n_B)^k}B_{n,k}\left(\frac{\partial n_B}{\partial(\mu_B/T)},\frac{\partial^2 n_B}{\partial(\mu_B/T)^2},..., \frac{\partial^{n-k+1} n_B}{\partial(\mu_B/T)^{n-k+1}}\right), 
}
where $B_{n,k}$ are partial exponential Bell polynomials. The l.h.s. of Eq.~(\ref{faa}), $\partial^nf/\partial(\mu_B/T)^n$, equals unity for $n=1$ and is zero for all $n>1$. Using $B_{n,1}\left(\frac{\partial n_B}{\partial(\mu_B/T)},\frac{\partial^2 n_B}{\partial(\mu_B/T)^2},..., \frac{\partial^{n} n_B}{\partial(\mu_B/T)^{n}}\right)=\frac{\partial^{n} n_B}{\partial(\mu_B/T)^{n}}$, Eq.~(\ref{faa}) can be presented in the form,
\eq{\label{faa2}
   \frac{\partial g}{\partial n_B}\frac{\partial^n n_B}{\partial (\mu_B/T)^n} = \frac{\partial^nf}{\partial(\mu_B/T)^n}-\sum_{k=2}^{n}\frac{\partial^kg}{\partial(n_B)^k}B_{n,k}\left(\frac{\partial n_B}{\partial(\mu_B/T)},\frac{\partial^2 n_B}{\partial(\mu_B/T)^2},..., \frac{\partial^{n-k+1} n_B}{\partial(\mu_B/T)^{n-k+1}}\right),
}
which links the $n$-th order derivative of $n_B$ with all of its lower order derivatives.
For $n=1$ we obtain the following from Eq.~(\ref{faa}):
\eq{\label{n1}
\frac{\partial n_B}{\partial(\mu_B/T)}=\left[\frac{\partial g}{\partial n_B}\right]^{-1}=n_B\left[\frac{1}{(1-b n_B)^2}-\frac{2 a n_B}{T}\right]^{-1}.
}
By substituting (\ref{n1}) in Eq.~(\ref{faa2}) one can calculate the second order derivative, $\partial^2 n_B/\partial(\mu_B/T)^2$. 
The procedure can then be applied iteratively to evaluate all derivatives of $n_B$ w.r.t $\mu_B/T$  up to a desired order.
The baryon number susceptibilities $\chi_{2m}$ are evaluated from Eq.~(\ref{chi2m-n}).

\end{widetext}

\bibliography{NuclearAnalyt}

\end{document}